\documentclass[fleqn,twoside]{article}
\usepackage[headings]{espcrc2}
\usepackage{epsfig,amsfonts}
\usepackage{graphicx}% Include figure files
\usepackage{dcolumn}% Align table columns on decimal point
\usepackage{bm}% bold math
\usepackage{amssymb}
\usepackage{amsfonts}
\usepackage{amsmath}
\usepackage{bbold}
\DeclareMathSymbol{\lesssim}      {\mathrel}{AMSa}{"2E}
\DeclareMathSymbol{\gtrsim}       {\mathrel}{AMSa}{"26}
\def\be{\begin{equation}}
\def\ee{\end{equation}}
\def\bc{\begin{center}}
\def\ec{\end{center}}
\def\bea{\begin{eqnarray}}
\def\eea{\end{eqnarray}}

\def\nn{\nonumber}
\newcommand{\afb}{A_{FB}}
\newcommand{\as}{\alpha_s}
\newcommand{\ord}{\mathcal{O}}

\newcommand{\sia}{\sigma_A}
\newcommand{\sis}{\sigma_S}

\title{Two-parton contribution to the heavy-quark $A_{FB}$ at NNLO QCD
in $e^+e^-$ collisions}

\author{R.~Bonciani\address[IFIC]{Departament de F\'{\i}sica Te\`orica, 
        IFIC, CSIC -- Universitat de 
        Val\`encia, E-46071 Val\`encia,
        Spain}%
        \thanks{This work was supported by MCYT (Spain) under 
	Grant FPA2004-00996, by Generalitat Valenciana (Grants GRUPOS03/013 
	and GV05/015).}}

\begin{document}

\begin{abstract}

At the next generation of linear colliders, forward-backward asymmetries will
be involved again in the precise determination of the neutral current couplings of heavy quarks
in inclusive heavy-quark production. The theoretical understanding of these 
observables at the level of NNLO radiative corrections is required. We review a 
recent calculation of the two-parton contribution to the heavy-quark $A_{FB}$ 
at NNLO QCD in $e^+e^-$ collisions. The results are valid for arbitrary values 
of the momentum transfer and non-vanishing heavy-quark mass $m_Q$.

\vspace{1pc}
\end{abstract}

\maketitle

%\section{Introduction}

The precision measurements of the forward-backward asymmetries in the production
of fermions at high-energy $e^+e^-$ colliders provide the determination of the
respective fermion neutral current couplings with a remarkable accuracy. For 
instance, the forward-backward asymmetry of $b$ quarks measured at the $Z$ 
resonance with a 1.7\% accuracy, led to a determination of 
$\sin^2 \theta_{W,eff}$ of the Standard Model with a relative precision of about 
1 per mille \cite{:2004qh,Abbaneo:1998xt}.

At a future linear $e^+ e^-$ collider \cite{tesla}, precision determinations of 
electroweak parameters will again involve forward-backward asymmetries. When such 
a collider will be operated at the $Z$ peak, accuracies of about 0.1\% may 
be reached for these observables \cite{Hawkings:1999ac,Erler:2000jg}. Moreover, 
the top quark asymmetry $A_{FB}^t$ will be experimentally accessible.
Therefore, it becomes important a theoretical understanding of these 
observables at the level of NNLO radiative corrections.

At present, the forward-backward asymmetry for the $c$ and $b$ quarks are known
at the level of NLO electroweak \cite{Bohm:1989pb,Bardin:1999yd,Freitas:2004mn}
and fully massive NLO QCD \cite{Jersak:1981sp,Arbuzov,Djouadi} corrections.
The NNLO QCD corrections were calculated in the limit of massless quarks in 
\cite{Altarelli,vanNeerven} and retaining logarithmically-enhanced terms 
of the type $\ln (Q/m_Q)$ in \cite{seymour}, where the calculation of $A_{FB}^b$
is done both with respect to the quark and the thrust axis.

In view of the future perspectives for the $b$- and $t$-quark asymmetries 
at a linear collider,  a computation of  the order $\as^2$ contributions to
$A_{FB}^Q$ for massive quarks $Q$ is clearly desirable. The NNLO QCD corrections
involve three classes of contributions: (1) the two-loop corrections to the 
decay of a vector boson into a heavy quark-antiquark pair; (2) the one-loop 
corrected matrix elements for the decay of a vector boson into a heavy 
quark-antiquark pair plus a gluon; (3) the tree level matrix elements for 
the decay of a vector boson into four partons, at least two of which being 
the heavy quark-antiquark pair.

The contributions (1) from the two-parton final state, and (2) plus
(3), i.e., those from the three- and four-parton final states, are separately
infrared-finite. The latter can be obtained along the lines of the calculations
of three-jet production involving heavy quarks 
\cite{Bernreuther:1997jn,Rodrigo:1997gy,Nason:1997tz,Bernreuther:2000zx}.   
However, a full computation of $A_{FB}^{3+4 \, parton}$ has not yet been done 
for massive quarks.

For what concerns the contribution (1), instead, an analytic expression valid
for non-vanishing heavy-quark mass $m_Q$ and arbitrary momentum transfer was
given in \cite{us0}. This expression lies on a previous calculation of the 
NNLO QCD corrections to the form factors for the vertices $\gamma Q \bar{Q}$ 
and $Z Q \bar{Q}$ \cite{us1,us2,us3,us4}, done using the Laporta algorithm 
\cite{Lap} for the reduction of the dimensionally-regularized scalar integrals 
to the set of master integrals and the differential equations technique 
\cite{DiffEq} for their calculation \cite{MIsus} in terms of harmonic 
polylogarithms \cite{HPLs} (all the calculations are done in FORM \cite{FORM}).

Forward-backward asymmetry for heavy quarks can be defined as the ratio of 
the ``antisymmetric'' and ``symmetric'' cross sections, $\sigma_A$ and
$\sigma_S = \sigma$,
\begin{align}
\afb = \frac{\sia}{\sis}\label{defafb},
\end{align}
where $\sigma_A$ and $\sigma_S$ are defined by:
\begin{align}
\hspace*{-7mm} \sia = & \int_0^1\frac{d\sigma}{d\cos\vartheta} d\cos\vartheta - 
\int_{-1}^0\frac{d\sigma}{d\cos\vartheta} d\cos\vartheta\label{defsia},\\
\hspace*{-7mm} \sis = & \int_{-1}^1\frac{d\sigma}{d\cos\vartheta} d\cos\vartheta \, .
\label{defsis}
\end{align}
$\vartheta$ is the angle between the incoming electron and the direction 
defining the forward hemisphere (in the $e^+ e^-$ center-of-mass frame), that
must be infrared- and collinear-safe in order that $\afb$ is computable
in perturbation theory. Common choices are the direction of flight of
$Q$ or the thrust axis direction. In the case of the $Q \bar Q$ contribution 
to $\afb$, $A^{(2p)}_{FB}$ (two-parton contribution), the result in the two cases is the same.

Expanding in series of $\as$ both $\sigma_A$ and $\sigma_S$, and retaining only
the contributions coming from the two-parton final state, we can write
$A^{(2p)}_{FB}$ as follows:
\begin{align}\label{afbent}
A^{(2p)}_{FB} =&
A_{FB,0}\;\left[1\;+\;A^{(2p)}_{1} \;+\;
A^{(2p)}_{2}\right] \, ,
\end{align}
where $A_{FB,0}$ is the forward-backward asymmetry at the Born level, 
and  $A^{(2p)}_{1}$ and $A^{(2p)}_{2}$ are the $\ord(\as)$ and  $\ord(\as^2)$ 
contributions normalized to $A_{FB,0}$. Their expressions in terms of the 
form factors can be found in \cite{us0}. Note that, although the single form 
factors, ultraviolet-renormalized, contain still infrared divergencies, the two-parton 
contributions to the forward-backward asymmetry $A^{(2p)}_{1}$ and 
$A^{(2p)}_{2}$ are infrared-finite.
This is due to the factorized structure of the infrared singularities at 
the level of the form factors. For massless multiloop amplitudes the
factorization of the infrared poles can be derived from exponentiation 
\cite{Catani:1998bh,Sterman:2002qn}. In the case of massive QCD, infrared 
factorization was established at the one-loop level in \cite{Catani:2000ef}, 
and the pole structure of the massive form factors calculated in 
\cite{us1,us2,us3,us4} suggests that it holds at higher orders as well.

\begin{figure}
\begin{center}
\epsfig{file=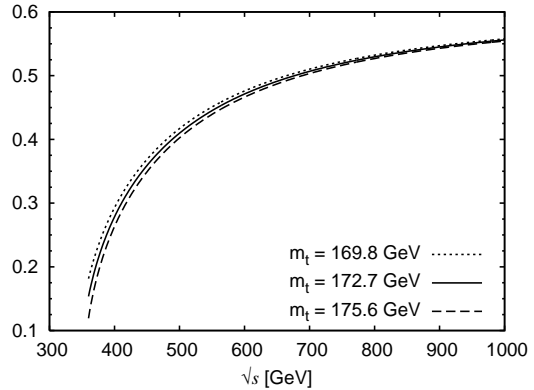, width=7.5cm, height=5.3cm}
\end{center}
\vspace*{-10mm}
\caption{Leading order asymmetry $A_{\mbox{\tiny{FB}},0}^{(t\bar t)}$
for three values of the top quark mass.}\label{fig:t0}
\vspace*{-3mm}
\end{figure}

\begin{figure}
\begin{center}
\epsfig{file=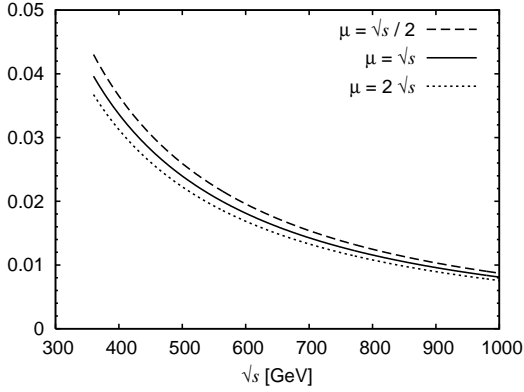, width=7.5cm, height=5.3cm}
\end{center}
\vspace*{-10mm}
\caption{Order $\as$ correction  $A_{1}^{(t \bar t)}$
for three values of the renormalization scale $\mu$, using
$m_t=172.7$ GeV.}\label{fig:t1ren}
\end{figure}

\begin{figure}
\begin{center}
\epsfig{file=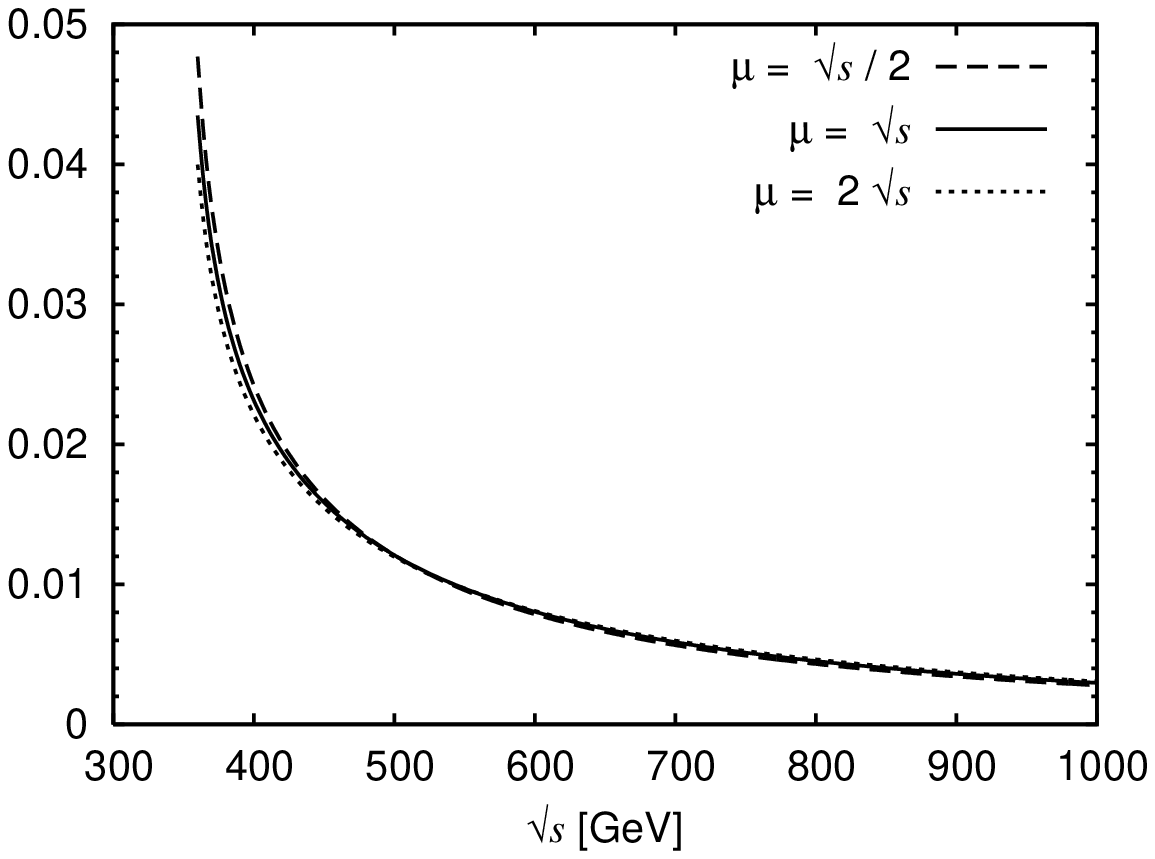, width=7.5cm, height=5.3cm}
\epsfig{file=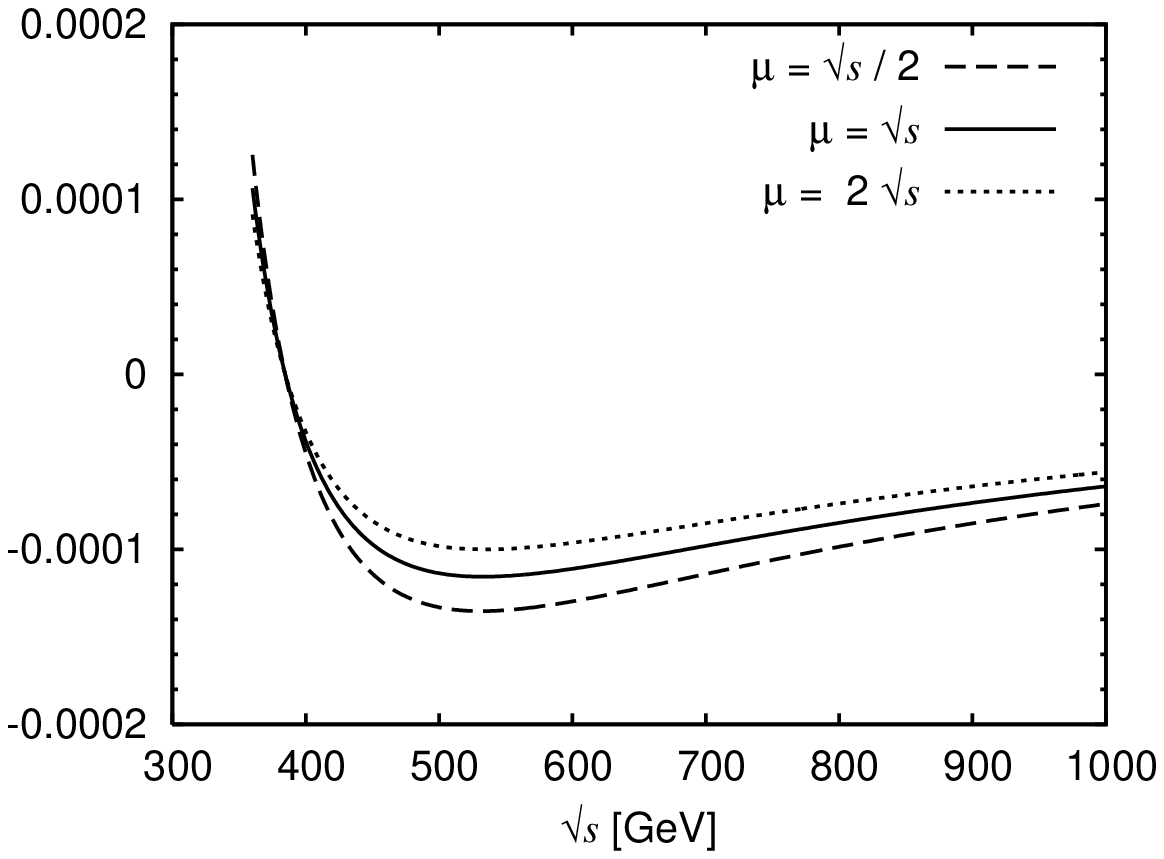, width=7.5cm, height=5.3cm}
\end{center}
\vspace*{-10mm}
\caption{Order $\as^2$ correction  $A_{2}^{(t\bar t,A)}$
(upper figure) and $A_{2}^{(t\bar t,B)}$ (lower figure)
for three values of the renormalization scale $\mu$, using
$m_t=172.7$ GeV.}\label{fig:t2aren}
\end{figure}

\begin{figure}
\begin{center}
\epsfig{file=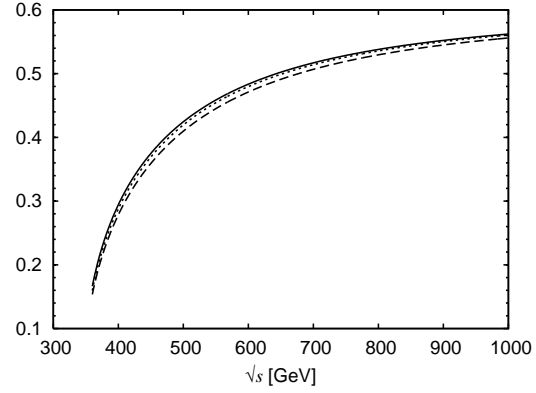, width=7.5cm, height=5.3cm}
\end{center}
\vspace*{-10mm}
\caption{Forward-backward asymmetry to lowest, first and second order
in $\as$ using $m_t=172.7$ GeV and
$\mu=\sqrt{s}$. $A_{FB,0}^{(t\bar t)}$ (dashed),
$A_{FB}^{(t\bar t)}(\as)$ (dotted), 
$A_{FB}^{(t\bar t)}(\as^2)$ (solid).}
\label{fig:tg012}
\vspace*{-5mm}
\end{figure}

In Fig.~\ref{fig:t0} $A_{FB,0}$ is shown in the case of production of a $t \bar{t}$ 
pair, for three different values of the 
mass of the top quark and for c.m. energy up to 1 TeV.
In Fig.~\ref{fig:t1ren} we give the order $\as$
correction, $A^{(2p)}_{1}$, for three different values of the renormalization
scale $\mu$. This correction ranges between 4\% near the threshold of 
production of the $t \bar{t}$ pair and less than 1\% at 1 TeV.
At the NNLO in QCD, $A^{(2p)}_{2}$ takes two kind of contributions called type 
A and type B contributions. Type A contributions are those where the $Z$ boson
couples directly to the external quark $Q$~\cite{us2}, while the triangle 
diagram contributions, summed over the  quark isodoublets of the three 
generations, are called type B~\cite{us3}. 
%(In the terminology of~\cite{seymour}
%these correspond to universal and non-universal corrections.)
$A_{2}^{(t\bar t,A)}$ and $A_{2}^{(t\bar t,B)}$ are shown in 
Fig.~\ref{fig:t2aren}, for three values of the renormalization scale.
Finally, in Fig.~\ref{fig:tg012} it is shown the forward-backward asymmetry 
to lowest, first and second order in $\as$, always in the case of top-quark
production, for $m_t=172.7$ GeV and $\mu=\sqrt{s}$.

In general the QCD corrections increase as far as the c.m. energy approaches
the threshold region, signaling that perturbation theory in $\as$ is no longer
applicable. At $\sqrt{s} =$ 400 GeV, $A_{1}^{(t\bar t)}$ is 
about 3.3\% while $A_{2}^{(t\bar t)}$ is about 2.4\%. 
Increasing the c.m. energy, the relative importance of the radiative corrections
decreases.
Moreover, we see that the QCD corrections do not depend strongly on the change
of the renormalization scale $\mu$.

\begin{figure}
\begin{center}
\epsfig{file=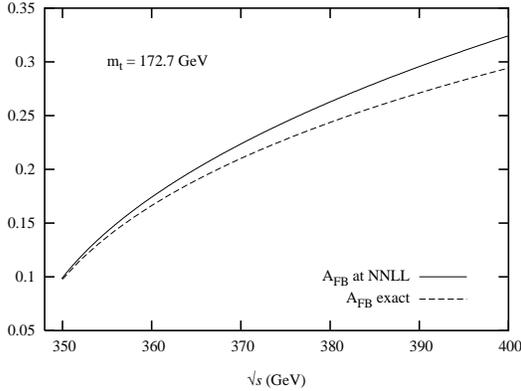, width=7.5cm, height=5.3cm}
\end{center}
\vspace*{-10mm}
\caption{The second order forward-backward asymmetry 
$A_{FB}^{(t\bar t)}(\as^2)$
near threshold: exact values (dashed) as given in
Fig.~\ref{fig:tg012} and the values obtained from the near-threshold
formula (solid), using $\mu=m_t=172.7$ GeV.}
\vspace*{-3mm}
\label{fig:trex}
\end{figure}

In \cite{us0,us5}, a detailed analysis of the behaviour of the exact formulas
in the energy region near the threshold of production of the $t \bar{t}$ pair 
is also done. For values of the quark velocity $\beta$ such that $\as \ll \beta
\ll 1$, the heavy-quark production cross section can be written as follows:
\bea
\! \! \! \! \sigma_{NNLO} \! \! & = & \! \! 
\sigma^{(2,0,\gamma)}_S \Bigl\{ 1 + \Delta^{(0,Ax)}
\nn\\
\! \! \! \! \! \! & &  
+ C_{F} \left( \frac{\alpha_{s}}{2 \pi} \right) \Delta^{(1,Ve)} \bigl( 1
+ \Delta^{(0,Ax)} \bigr) 
\nn\\
\! \! \! \! \! \! & &  
+ C_{F} \left( \frac{\alpha_{s}}{2 \pi} \right) 
{\Delta}^{(1,Ax)}
\nn\\
\! \! \! \! \! \! & &  
+ C_{F} \left( \frac{\alpha_{s}}{2 \pi} \right)^2  \Delta^{(2,Ve)}
\bigl( 1 + \Delta^{(0,Ax)} \bigr)
\nn\\
\! \! \! \! \! \! & &  
+ C_{F} \left( \frac{\alpha_{s}}{2 \pi} \right)^2 
\Delta^{(2 ,Ax)} \Bigr\} \, ,
\label{CS}
\eea
where the various terms are given in Laurent series of $\beta$ (Coulomb poles
in $1/\beta$ are present). At order $\beta^0$ the cross section in 
Eq.~(\ref{CS}) is infrared finite. The analytic expressions of $\Delta^{(i,Ve)}$
and $\Delta^{(0,Ax)}$ agree with the results in 
\cite{Beneke:1999qg,Hoang:2001mm,Czarnecki:1997vz,Hoang:1997sj,AX}, while
in \cite{us0,us5} the term 
\be
\hspace*{-3mm} \Delta^{(2,Ax)} \! = \! 
\frac{64 \zeta(2) m_Q^4 (a_Q^Z)^2 \! \! \left[ (v_e^Z)^2 \! + \! (a_e^Z)^2 \right]}{
( v_Q^{\gamma} v_e^{\gamma} )^2 (4m_Q^2 - m_Z^2)^2} C_F 
\label{Z2} 
\ee
is also given. These results, however, can not be used to extract the matching 
coefficients at two loops between QCD and NRQCD \cite{Kniehl:2006qw}.

Using Eq.~(\ref{CS}) and an analogous expression for the antisymmetric cross
section, we can compute the forward-backward asymmetry near the threshold, that,
at this order in $\beta$, is equal to the complete forward-backward asymmetry 
$A_{FB}^{(t\bar t)}$. In Fig.~\ref{fig:trex} we show the comparison 
between the exact second order forward-backward asymmetry 
$A_{\mbox{\tiny{FB}}}^{(t\bar t)}(\as^2)$  as given in
Fig.~\ref{fig:tg012} and the values obtained from the near-threshold
formula. For $\sqrt{s} \lesssim$ 360 GeV
corresponding to $\beta \lesssim 0.3$ the deviation of the threshold from
the respective exact value is less than 5\%.

\vspace*{3mm}

\noindent {\bf Acknowledgments} 

The author wishes to thank W.~Bernreuther, T.~Gehrmann, R.~Heinesch,
T.~Leineweber, P.~Mastrolia and E.~Remiddi for their collaboration.

\end{document}